# The Optimized path for the public transportation of Incheon in South Korea

line 1: 1st Soroor Malekmohammadi faradounbeh
line 2: *Department of Game Mobile Engineering*
line 3: *Keimyung University*
line 4: Daegu, Korea
line 5: soroormalekmohamadi@gmail.com

line 1: 2nd Hongle Li
line 2: *Department of Game Mobile Engineering*
line 3: *Keimyung University*
line 4: Daegu, Korea
line 5: pipi9530@gmail.com

line 1: 3th Mangkyu Kang
line 2: *Department of Game Mobile Engineering*
line 3: *Keimyung University*
line 4: Daegu, Korea
line 5: audrb256@gmail.com

line 1: 4th Choongjae Im
line 2: *Department of Game Mobile Engineering*
line 3: *Keimyung University*
line 4: Daegu, Korea
line 5: 87dooly@naver.com

*Abstract*— path-finding is one of the most popular subjects in the field of computer science. Pathfinding strategies determine a path from a given coordinate to another. The focus of this paper is on finding the optimal path for the bus transportation system based on passenger demand. This study is based on bus stations in Incheon, South Korea, and we show that our modified A* algorithm performs better than other basic pathfinding algorithms such as the Genetic and Dijkstra. Our proposed approach can find the shortest path in real-time even for large amounts of data(points).

## I. INTRODUCTION

Public transportation system is one of the services that people often use and finding the path is one of the main technologies. Finding the best routes while planning a trip or choosing a route is a multi-objective challenge for transit systems. Travelers look for routes that require the least amount of time, money, transfers, and other factors. However, in a genuine transit network, goals are frequently in conflict, forcing users to choose between their goals [1]. Also, finding the optimal path, in addition to saving passenger time and travel distance, helps to reducing the air pollution (like Greenhouse Gas emission) and reduce fuel consumption by eliminating extra routes. Other benefits attributable to public transport include less congestion, preservation of open space and the reduction of urban sprawl [2-5].

Pathfinding has a history dating back to the 19th century and is considered to be a classic graph problem. It gained prominence in the early 1950s in the context of alternate routing; that is, finding the second-shortest path if the shortest path is blocked. In 1956, Edsger Dijkstra created the best-known of these algorithms [6]. The discovery and mapping of an optimal path between two points on a plane are referred to as the path-finding problem. These kinds of systems take a start point and a destination into account, after which they identify a succession of points that collectively make up a path to the target. A pre-computed data structure is typically used by the AI (Artificial Intelligence) pathfinder to direct movement.

The existing algorithms that address this issue are largely static and significantly reliant on preexisting environmental knowledge. They also demand a predictable environment. However, in practical applications of the path-finding issue, the environment is frequently unpredictable, previously unknown, and has multiple competing goals. The aforementioned algorithms are ineffective in such situations. In transportation system for specific aims, like shuttle buses, they need to find the new path based on the new situation in real-time instead to use the static path all the time, thus the method should be able to react and change in real-time to any dynamic changes that may occur in the path. The two main components for basic real-time pathfinding are (1) traveling to-wards a specified goal and (2) avoiding dynamic and static obstacles that may litter the path to this goal.

Finding the shortest path to save time and increase efficiency has always been a very important and significant point in this regard, and many algorithms are provided for it [7]. The shortest path algorithm determines the low cost and shortest route between two nodes. It works in real time, making it helpful for user interactions and dynamic workflows, but always we have to import two nodes as a start and end point to this type of algorithms to find the shortest path. Therefore, in original way, they are not too efficient for transit systems, due to need to cover some specific points(stations). Another problem is finding the optimized order of stations and rearrange them to have the lowest cost (the cost can be distance, time, and any other weight or combine of them).

In this paper, we present the approach to find the optimized path for public transportation that in addition to finding the shortest path, the final founded path will have some other conditions, including the coverage of all stations that have passengers. In general, the pathfinder algorithms find the path between two specific points, but in this case, only have the starting point and some other points that should be covered.

This paper is organized as follows. Section 2 discusses the previously proposed algorithms. Section 3 describes and evaluates our method. Section 4 concludes..

## II. OVERVIEW OF ALGORITHMS

In order to determine Dijkstra's shortest path algorithm is first finding the lowest-weight relationship between the start node and all directly connected nodes [8]. The node that is "closest" is chosen by keeping track of those weights. The calculation is then repeated, this time as a total cumulative starting from the start node. Until it reaches the destination node, the algorithm keeps doing this, evaluating a "wave" of cumulative weights, and always choosing the lowest weighted cumulative path to move along.

Graph search algorithms are the foundation upon which pathfinding algorithms are built. These algorithms look for connections between nodes, starting at a single node and moving via relationships until they reach their target. For applications like logistics planning, least-cost call or IP routing, and gaming simulation, these algorithms are used to find the best paths through a graph.

In either general discovery or explicit search, graph search algorithms investigate a graph. These methods carve pathways across the graph, but their computational efficiency is not expected. Some of the common pathfinding (shortest pathfinding) algorithm as the searching graph algorithm are Breadth-First Search (BFS), Depth First Search (DFS), Floyd warshall, Bellman fold, Dijkstra, A*, K shortest path. Also, the common previous methods empowered by neural networks, machine learning algorithms, or other heuristics/meta-heuristic methods [9-12]. These are frequently essential to traversing a graph and also serve as the initial step in many other forms of analysis.

A* algorithm is one of the fastest and most popular pathfinding algorithms. Algorithm A * repeatedly is searching the most promising (depends on the goal function) un-discovered locations. When a location is explored, if the target is that location, the algorithm ends. Otherwise, all the neighbors of that location will be kept on a list for further searching [ 13,14].

The shortest bus routes between the user-specified present and destination are one of the most crucial pieces of information to be provided to consumers of public transportation [15]. The challenging task of determining the shortest pathways on very large road networks is decreased if the transit node routing precomputes distance tables for significant transit nodes and all pertinent connections between the remaining nodes and the transit nodes [16]. It is very helpful to reduce calculation time and the number of real-time arrival requests to the transportation agency as well as the server bottleneck in calculations by using a pre-computed lookup table of potential routes between the origin station of each bus route and the terminus of any other bus route using transfer points [16].

Some previous research has been proposed methods of path planning (pathfinding) and routing on public transportation [5-11]. Nevertheless, it is necessary to modify the algorithm to adjust with the real case [17].

## III. OUR APPROACH

The idea is to do Depth first traversal of given directed graph to search all the exist path. Also use the A* algorithm to find the shortest path based on the costs of each path. beginning the traverse at the source,

To prevent a traverse cycle, keep saving the visited vertices in an array called "path []"; once you reach the destination vertex, report the contents of "path []" as visited. Also, the "Cover" method receives the stations that have passengers (after checking the possibility of bus travel). The task of this method is to ensure that all stations with passengers are covered in the proposed path. And also, we used the Manhattan distance function [18] as a heuristic function for our A* algorithm.

$$\text{Distance} = |x_1 - x_2| + |y_1 - y_2| \qquad (1)$$

All the data that be needed information about the all the location of stations and also information about path between them for example one-way or two-way, cost of each path, and so on.

The input of this algorithm is just the starting point and destination point of each passenger. And follow this flow chart:

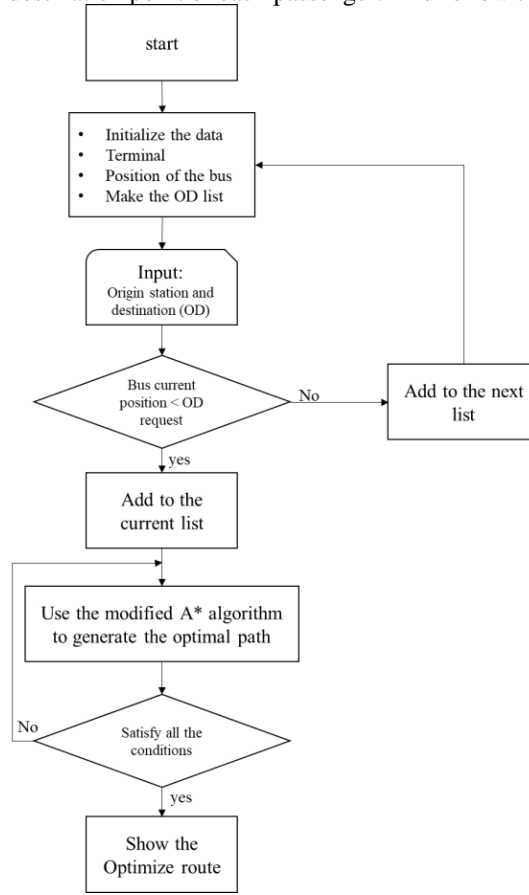

Fig. 1. Flowchart of our approach

In the worst situation of an unlimited search space, the number of nodes expanded is exponential in the depth of the solution, which determines the A* time complexity: O(n log n), where n is the number of nodes (vertices) [19], therefore use it in a large graph will not be optimal. Also, our purpose to find the shortest path with cover all the stations that have requested by the passengers. The basic pathfinding algorithms need two points: start point and endpoint to find the shortest path, so these algorithms cannot be used directly for our problem, but they will be useful to make a new graph. we use A* algorithm to find the shortest path between each pair stations. So, this means that it is possible to create a smaller graph (undirected, weighted, and fully connected) that instead of the first graph. In the new graph only exist bus stations (as vertices) and use the shortest distance for each edge.

The advantage of this method is that we can easily change the new graph to even a smaller graph (because it is fully connected and eliminating some vertices is not a problem) and a smaller graph means a higher calculation speed (less time consuming) and less memory usage. All these actions useful to make the real time system, and the figure 2 shows how change the big graph to the little one that the system needed to calculation.

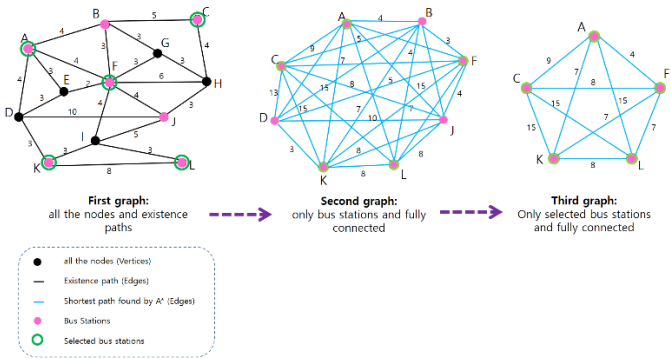

Fig. 2. Example of convert the big graph to the graph in question

After this processing to find an optimized path that covers all selected bus stations, we can use the final new graph and find it with fewer searches rather than the first graph. Another condition that our approach satisfied is to visit the O (Origin) point before the D (Destination) point of passengers, which prevents the bus to visit the stations without any passengers or visit the passenger's destination's station before they take in the bus.

in the following, we listed all the conditions that our approach satisfied them:

Cover all bus stations

Find the shortest path

Only the starting point is specified and we have no end point, that mean based on the passenger's request end point can be changeable.

The final path has not a cycle

Optimal order(sequence) to visiting the stations

Given that by doing the preprocessing, at last, we have a fully connected smaller graph and just need to pass all the vertices (selected bus stops) only once. also, we use a greedy method to continue calculations that make it very fast, these steps are described in pseudocode as shown in Figure 3. Start with the starting point (vertex) and to find the next vertices, find the edge that costs less to reach the new vertex and mark the found vertex as the visited node, thus visiting all the vertices and keep doing this until all the vertices will be marked as visited.

```
Algorithm : Find the Optimized Path (Order)
  Result: order of the stations
  initialization;
  n = selected stations;
  make the nxn array(matrix) that fills it by the distance-connection;
  if any two indexes haven't any connection the value is equal to infinity(like as 1 and 1);
  current-point = start-point;
  while all the selected stations visit do
     instructions;
     if current-point not visited then
        select the minimum value from current-point-th row of matrix;
        mark current-point as visited point;
        change all the value of this index in other rows to infinity;
        current-point = index of minimum value;
     end
     add index of visited stations to make a path;
  end
```

Fig. 3. Pseudocode of Optimal path finding algorithm.

IV. ANALYSIS AND EVALUATION

The case study of this paper is on all of the bus stations in Incheon city - south Korea, that contains the 191497 nodes where 272 of which are bus stations. We use the latitude and longitude to find the position of all the bus stations and use the .NET platform and OpenStreetMap to show the result. we also use the gmap library to upload the map in our project.

Our modified A * algorithm is compared with the other two methods in the algorithm time-consuming and the distance of path found. These two algorithms are: 1) Dijkstra as a basic method for pathfinder and 2) Genetic algorithm, have a high-speed calculation and also find the shortest path.

The scheme of our system with two found example routes is shown in figure 4, all the found routes are satisfied the conditions.

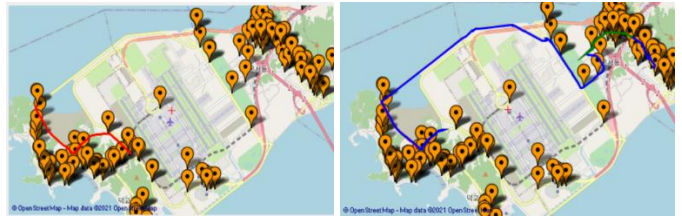

Fig. 4 Two examples of pathfinding on OSM and .NET platform, all the Incheon bus stops

As it shown in table I the calculation time and the distance of the path found both of them are lower than two other algorithms.

TABLE I. COMPARE THREE ALGORITHMS RESULTS TO FIND THE SHORTEST PATH BETWEEN TWO POINTS.

| Algorithm | Distance(m) of shortest path | Time calculation (ms) |
|---|---|---|
| Dijkstra | 712433.1118 | 19124 |
| Genetic | 6026.1744 | 84 |
| Modify A* (our approach) | 6028.1744 | 6 |

In Table II, we show the results of our approach and also Genetic for more than two bus stops that passengers request and consider the starting point as the first passenger request. As shown in the table, the more bus stops and requests, the performance of the genetic algorithm decrease, but our proposed method can still perform real-time path-finding and find the shortest possible path.

TABLE II. COMPARE OUR APPROACH WITH THE GENETIC ALGORITHM TO FIND THE OPTIMAL PATH BASED ON PASSENGER'S DEMAND

| Test case | Passenger's request list | Our approach's result Distance(m) / time (ms) | Genetic result Distance(m) / time (ms) |
|---|---|---|---|
| 1 | 254 → 66, 257 → 59 | 254→257→66→59 26964.0372 \| 46 | 254→257→66→59 29402.6884 \| 119 |
| 2 | 74 → 257, 65 → 30, 255 → 58 | 74→255→257→65→58→30 39310.4194 \| 52 | 74→65→58→30→255→257 49884.3667 \| 7215 |
| 3 | 255 → 96, 76 → 62, 258 → 58, 72 → 65, 337 → 47, 202 → 12, 154 → 194 | 255→258→76→72→96→65→62→58→154→47→202→194→12→337 69723.0433 \| 61 | 255→258→72→96→12→194→202→47→337→154→58→62→65→76 723579.37 \| 14258 |

## V. CONCLUSIONS

In this approach, with the help of A* algorithm and preprocessing on data, it is possible to find the shortest path in real-time even on large data. As a consequence, to find the high-performance shortest path based on the passengers' request (on demand), has the fastest on real road networks.

Our proposed method finds the optimal route for the bus on-demand in the shortest time. In addition to being the shortest route, this optimal route includes various conditions such as covering all the requested stations, visiting the origin stations before the destination, without cycling, and just only having one terminal.

And the results of comparison at the time of execution as well as the distance found, with two algorithms, Dijkstra and Genetic showed that our method has less execution time and finds a more efficient path with less distance.

This study also provides a general comparison in time and space between the famous shortest path finder algorithms.

In future studies, we intend to improve the demand list creation algorithm and consider more conditions in creating it as well as generating the route, using other factors like waiting time, traffic status, and so on, to make a more optimal shortest path in the rural and urban area. Also, using the Big Data (one or more years collected data of bus transportations in a specific area) can make the optimal static bus lines and order of stations in the urban areas for the routine bus transportation. also, other elements could be added in future works, such as using multiple vehicle types which involve different capacities and speeds or using more than one bus with different terminals.